\documentclass[12pt,amssymb,amsmath,tightenlines]{article}
\usepackage{amsmath, amssymb, amsthm, latexsym, mathrsfs, float}
\usepackage[pdftex]{graphicx}

\newcommand{\half}{\mbox{$\textstyle \frac{1}{2}$}}

\newcommand{\rd}{\mbox{$\rm d$}}
\newcommand{\PR}{\mathbb{P}}

\newcommand{\E}{\mathbb{E}}
\newcommand{\B}{\mathbb{B}}
\newcommand{\F}{\mathcal{F}}

\newcommand{\nn}{\nonumber}
\newcommand{\indi}[1]{1\hspace{-.09cm}\textup{\textrm{l}}}

\theoremstyle{definition}

\numberwithin{equation}{section}
\title{\bf Security Pricing with Information-Sensitive Discounting}
\begin{document}
\author{Andrea Macrina$^{\ast \dagger}$ and Priyanka A. Parbhoo$^{\ddagger}$}
\date{}
\maketitle
\vspace{-0.5cm}
\begin{center}
$^{\ast}$Department of Mathematics, King's College London, London, UK\\
$^{\dagger}$Institute of Economic Research, Kyoto University, Kyoto, Japan\\
$^{\ddagger}$School of Computational and Applied Mathematics,
\\University of the Witwatersrand, Johannesburg, South Africa
\end{center}
\vspace{0.25cm}
\begin{abstract}
In this paper incomplete-information models are developed for the
pricing of securities in a stochastic interest rate setting. In
particular we consider credit-risky assets that may include random
recovery upon default. The market filtration is generated by a
collection of information processes associated with economic
factors, on which interest rates depend, and information processes
associated with market factors used to model the cash flows of the
securities. We use information-sensitive pricing kernels to give
rise to stochastic interest rates. Semi-analytical expressions for
the price of credit-risky bonds are derived, and a number of
recovery models are constructed which take into account the
perceived state of the economy at the time of default. The price of a
European-style call bond option is deduced, and it is shown how
examples of hybrid securities, like inflation-linked credit-risky
bonds, can be valued. Finally, a cumulative information process is
employed to develop pricing kernels that respond to the amount of
aggregate debt of an economy.
\\
\begin{center}
3 June 2010 \vspace{.375cm}
\\
{\bf E-mail:} andrea.macrina@kcl.ac.uk, parbhoop@cam.wits.ac.za\\
\end{center}
\end{abstract}

\section{Introduction}
The information-based framework developed by Brody \emph{et
al}.~(2007, 2008a) is a method to price assets based on incomplete
information available to market participants about the cash flows of
traded assets. In this approach the value of a number of different
types of assets can be derived by modelling the random cash flows
defining the asset, and by explicitly constructing the market
filtration that is generated by the incomplete information about
independent market factors that build the cash flows. This principle
has been used in Brody \emph{et al}.~(2007) to derive the price
processes of credit-risky securities, in Brody \emph{et al}.~(2008a)
to value equity-type assets with various dividend structures, in
Brody \emph{et al}.~(2008b) to price insurance and reinsurance
products, and in Brody \emph{et al}.~(2009) to price assets in a
market with asymmetric information. However, for simplicity, in this
framework it is typically assumed that interest rates are
deterministic.

One of the earliest generalizations of the models developed in Brody
\emph{et al}.~(2007) to include stochastic interest rates can be
found in Rutkowski \& Yu (2007). Here, it is assumed that the
filtration is generated jointly by the information processes
associated with the future random cash flows of a defaultable bond
and by an independent Brownian motion that drives the stochastic
discount factor.

Pricing kernel models for interest rates have been studied in
Flesaker \& Hughston (1996), Hunt \& Kennedy (2004) and Rogers
(1997), among others. In such models, the price $P_{tT}$ at time $t$
of a sovereign bond with maturity $T$ and unit payoff, is given by
the formula
\begin{equation}\label{bpformula}
P_{tT} = \frac{\mathbb{E}^\mathbb{P}[\pi_T\,|\,\F_t]}{\pi_t},
\end{equation}
where $\{\pi_t\}_{t\geq 0}$ is the $\{\F_t\}$-adapted pricing kernel
process and $\PR$ denotes the real probability measure. Given the
filtration $\{\F_t\}_{t\geq 0}$, arbitrage-free interest rate models
can be obtained by specifying the dynamics of the pricing kernel. In
particular, term structure models with positive interest rates are
generated by requiring that $\{\pi_t\}$ is a positive
supermartingale. A more recent approach to constructing interest
rate models in an information-based setting, presented in Hughston
\& Macrina (2009), develops the notion of an information-sensitive
pricing kernel. The pricing kernel is modelled by a function of time
and information processes that are observed by market participants
and that over time reveal genuine information about economic factors
at a certain rate. In order to obtain positive interest rate models,
this function must be chosen so that the pricing kernel has the
supermartingale property. A scheme for generating appropriate
functions to construct such pricing kernels in an information-based
approach is considered in Akahori \& Macrina (2010). Incomplete
information about economic factors that is available to investors is
modelled in Akahori \& Macrina (2010) by using time-inhomogeneous
Markov processes. The Brownian bridge information process considered
in Hughston \& Macrina (2009) and, more generally, the subclass of
the continuous L\'evy random bridges, recently introduced in Hoyle
\emph{et al}.~(2009), are examples of time-inhomogeneous Markov
processes.

In this paper we describe how credit-risky securities can be priced
within the framework considered in Brody \emph{et al.}~(2007) while
including a stochastic discount factor by use of
information-sensitive pricing kernels. To this end, we proceed in
Section 2 to recap briefly the theory for the pricing of
fixed-income securities in an information-based framework described
in Hughston \& Macrina (2009). In Section 3 we recall the result in
Akahori \& Macrina (2010) that can be used to obtain the explicit
dynamics of the pricing kernel by use of so-called ``weighted heat
kernels'' with time-inhomogeneous Markov processes. In Section 4, we
derive the price process of a defaultable discount bond and compute
the yield spreads between digital bonds and sovereign bonds. Section
5 considers a number of random recovery models for defaultable
bonds, and in the following section we derive a semi-analytical
formula for the price of a European option on a credit-risky bond.
In Section 7 we demonstrate how to price credit-inflation securities
as an example of a hybrid structure. We investigate the valuation of
credit-risky coupon bonds in Section 8 and conclude by considering a
pricing kernel that reacts to the level of debt accumulated in a
country over a finite period of time.
\section{Information-sensitive pricing kernels}
We define the probability space $(\Omega,\mathcal{F},\{\F_t\}_{t\geq 0}, \PR)$, where $\PR$ denotes the real probability
measure. We fix two dates $T$ and $U$, where $T < U$, and introduce
a macroeconomic random variable $X_U$, the value of which is
revealed at time $U$. Noisy information about the economic factor
available to market participants is modelled by the information
process $\{\xi_{tU}\}_{0\leq t\leq U}$ given by
\begin{equation}
\xi_{tU} = \sigma\,t\,X_U + \beta_{tU}.
\end{equation}
Here the parameter $\sigma$ represents the information flow rate at
which the true value of $X_U$ is revealed as time progresses, and
the noise component $\{\beta_{tU}\}_{0\leq t\leq U}$ is a Brownian
bridge that is taken to be independent of $X_U$. We assume that the
market filtration $\{\F_t\}_{t\geq 0}$ is generated by
$\{\xi_{tU}\}$, and note that it is shown in, e.g., Brody \emph{et al}.~(2007) that
$\{\xi_{tU}\}$ is a Markov process with respect to its natural
filtration. We consider pricing kernels $\{\pi_t\}$ that are of the
form
\begin{equation}\label{pk in hm}
\pi_t = M_{t}\,f(t,\xi_{tU}),
\end{equation}
where $\{M_t\}_{0\leq\, t < U}$ is the density martingale associated
with a change of measure from $\mathbb{P}$ to the so-called ``bridge
measure" $\mathbb{B}$ under which the information process has the law of a
Brownian bridge. It is proven in Brody \emph{et al}.~(2007), that $\{M_t\}$
satisfies the differential equation
\begin{equation}\label{martingale}
\rd M_t
=-\,\sigma\frac{U}{U-t}\,\mathbb{E}^\mathbb{P}[X_U\,|\,\xi_{tU}]\,M_t\,\rd
W_t,
\end{equation}
where $\{W_t\}_{0\leq\, t < U}$ is an $(\{\F_t\}, \mathbb{P})$-Brownian motion given by
\begin{equation}
W_t = \xi_{tU} + \int_0^t\frac{1}{U-s}\,\xi_{sU}\,\rd
s-\sigma U\int_0^t\frac{1}{U-s}\,\mathbb{E}^\mathbb{P}[X_U\,\vert\,\xi_{sU}]\,\rd
s.
\end{equation}
By applying Bayes change-of-measure formula to equation
(\ref{bpformula}), we can express the price $P_{tT}$ at time $t$ of
a sovereign discount bond with maturity $T$ by
\begin{equation}
P_{tT} =
\frac{\mathbb{E}^{\mathbb{B}}[f(T,\xi_{TU})\,|\,\xi_{tU}]}{f(t,\xi_{tU})}.
\end{equation}
Next we introduce the random variable $Y_{tT}$ defined by
\begin{equation}
Y_{tT} = \xi_{TU}-\frac{U-T}{U-t}\xi_{tU},
\end{equation}
and observe that under the measure $\B$, $Y_{tT}$ is a Gaussian
random variable with zero mean and variance given by
\begin{equation}\label{v}
\nu^2_{tT}=\frac{(T-t)(U-T)}{U-t}.
\end{equation}
It can be verified that $Y_{tT}$ is independent of $\xi_{tU}$ under
$\mathbb{B}$, see Hughston \& Macrina (2009). Next, we introduce a
Gaussian random variable $Y$, with zero mean and unit variance; this
allows us to write $Y_{tT} = \nu_{tT}Y$. Since $\xi_{tU}$ is
$\F_t$-measurable and $Y$ is independent of $\xi_{tU}$, we can
express the price of a sovereign bond by the following Gaussian
integral:
\begin{equation}\label{hm}
P_{tT} = \frac{1}{f(t,\xi_{tU})}\int_{-\infty}^\infty f\left(T, \nu_{tT}y + \frac{U-T}{U-t}\xi_{tU}\right)\frac{1}{\sqrt{2\pi}}\exp{\left(-\tfrac{1}{2}y^2\right)}\,\rd y.
\end{equation}
Interest rate models of various types can therefore be constructed
in this framework by specifying the function $f(t,x)$. However,
pricing kernels constructed by the relation (\ref{pk in hm}) are not
automatically $(\{\mathcal{F}_t\},\mathbb{P})$-supermartingales. In
particular, to guarantee positive interest rates, it is a
requirement that the function $f(t,x)$ satisfies the following
differential inequality, see Hughston \& Macrina (2009):
\begin{equation}\label{ineq}
\frac{x}{U-t}\,\frac{\partial}{\partial x}\,f(t,x)-
\frac{1}{2}\,\frac{\partial^2}{\partial^2
x}\,f(t,x)-\frac{\partial}{\partial t}\,f(t,x)>0.
\end{equation}
We emphasize that finding a function which satisfies relation
(\ref{ineq}) is equivalent to finding a process
$\{f(t,\xi_{tU})\}_{0\leq \,t <U}$ that is a positive
supermartingale under the measure $\mathbb{B}$. Hence the pricing
kernel $\{\pi_t\}_{0\leq \,t <U}$ is a positive
$(\{\mathcal{F}_t\},\mathbb{P})$-supermartingale since
\begin{align}
\mathbb{E}^\mathbb{P}[\pi_T\,|\,\mathcal{F}_t] = M_t\,\mathbb{E}^{\mathbb{B}}[f(T,\xi_{TU})\,|\,\xi_{tU}] \leq M_t\,f(t,\xi_{tU}) = \pi_t.
\end{align}
We now proceed to construct such positive
$(\{\mathcal{F}_t\},\mathbb{B})$-supermartingales using a technique
known as the ``weighted heat kernel approach'', presented in Akahori
\emph{et al}.~(2009) and adapted for time-inhomogeneous Markov
processes in Akahori \& Macrina (2010).
\section{Weighted heat kernel models}
We consider the filtered probability space
$(\Omega,\F,\{\F_t\},\PR)$ where the filtration $\{\F_t\}_{t\geq 0}$
is generated by the information process $\{\xi_{tU}\}$. We recall
that the martingale $\{M_t\}$ satisfying equation
(\ref{martingale}), induces a change of measure from $\mathbb{P}$ to
the bridge measure $\mathbb{B}$, and that the information process
$\{\xi_{tU}\}$ is a Brownian bridge under $\mathbb{B}$. The Brownian
bridge is a time-inhomogeneous Markov process with respect to its
own filtration. Let $w:\mathbb{R}^+_0 \times \mathbb{R}^+_0
\rightarrow \mathbb{R}^+$ be a weight function that satisfies
\begin{equation}\label{weight_ineq}
w(t,u-s) \leq w(t-s,u)
\end{equation}
for arbitrary $t,u\in \mathbb{R}^+_0$ and $s\leq t\wedge u$. Then,
for $t<U$ and a positive integrable function $F(x)$, the process
$\{f(t,\xi_{tU})\}$ given by
\begin{equation}\label{weight}
f(t,\xi_{tU}) = \int_0^{U-t} \mathbb{E}^{\mathbb{B}}[F(\xi_{t+u,U})\,|\,\xi_{tU}]\,w(t,u)\,\rd u
\end{equation}
is a positive supermartingale.

The proof of this result goes as follows. For $f(t,x)$ an integrable
function, the process $\{f(t,\xi_{tU})\}$ is a supermartingale for
$0\leq s\leq t<U$ if
\begin{equation}
\mathbb{E}^{\mathbb{B}}[f(t,\xi_{tU})\,|\,\xi_{sU}] \leq f(s,\xi_{sU})
\end{equation}
is satisfied. We define the process $\{p(t,u,\xi_{tU})\}$ by
\begin{equation}
p(t,u,\xi_{tU})=\E^{\B}\left[F(\xi_{t+u,U})\vert\,\xi_{tU}\right],
\end{equation} where $0\leq u \leq U-t$. Then we have:
\begin{eqnarray}
\mathbb{E}^{\mathbb{B}}[f(t,\xi_{tU})\,|\,\xi_{sU}] &=& \int_0^{U-t}\mathbb{E}^{\mathbb{B}}[p(t,u,\xi_{tU})\,|\,\xi_{sU}]\,w(t,u)\,\rd u\nn\\
&=& \int_0^{U-t}p(s, u+t-s, \xi_{sU})\,w(t,u)\,\rd u\nn\\
&=& \int_{t-s}^{U-s} p(s,v,\xi_{sU})\,w(t,v-t+s)\,\rd v.
\end{eqnarray}
Here we have used the tower rule of conditional expectation and the
Markov property of $\{\xi_{tU}\}$. Next we make use of the relation
(\ref{weight_ineq}) to obtain
\begin{eqnarray}
\mathbb{E}^{\mathbb{B}}[f(t,\xi_{tU})\,|\,\xi_{sU}] &\leq& \int_{t-s}^{U-s}p(s,v,\xi_{sU})\,w(t-(t-s),v)\,\rd v\nn\\
&\leq& \int_{0}^{U-s} p(s,v,\xi_{sU})\,w(s,v)\,\rd v = f(s,\xi_{sU}).
\end{eqnarray}
Thus, $\{f(t,\xi_{tU})\}$ is a positive $(
\{\mathcal{F}_t\},\mathbb{B})$-supermartingale if $F(x)$ is positive.

The method based on equation (\ref{weight}) provides one with a
convenient way to generate positive pricing kernels driven by the
information process $\{\xi_{tU}\}$. These models can be used to
generate information-sensitive dynamics of positive interest rates.
In particular, the functions $f(t,x)$ underlying such interest rate
models satisfy inequality (\ref{ineq}).
\section{Credit-risky discount bonds}
We introduce two dates $T$ and $U$, where $T<U$, and attach two
independent factors $X_T$ and $X_{U}$ to these dates respectively.
We assume that $X_T$ is a discrete random variable that takes values
in $\{x_0, x_1,\ldots, x_n\}$ with a priori probabilities $\{p_0,
p_1, \ldots, p_n\}$, where $1\ge x_n > x_{n-1} > \ldots > x_1 >
x_0\ge 0$. We take $X_T$ to be the random variable by which the
future payoff of a credit-risky bond issued by a firm is modelled. The
second random variable $X_{U}$ is assumed to be continuous and
represents a macroeconomic factor. For instance, one might consider
the GDP level at time $U$ of an economy in which the bond is issued.
With the two $X$-factors, we associate the independent information processes
$\{\xi_{tT}\}_{0\leq t\leq T}$ and $\{\xi_{tU}\}_{0\leq t\leq U}$
given by
\begin{align}
&\xi_{tU} =\sigma_1\,t\,X_U+\beta_{tU},&
\xi_{tT} =\sigma_2\,t\,X_{T} + \beta_{tT}.
\end{align}
The market filtration $\{\mathcal{F}_t\}$ is generated by both
information processes $\{\xi_{tT}\}$ and $\{\xi_{tU}\}$. The price
$B_{tT}$ at $t\leq T$ of a defaultable discount bond with payoff $H_T$
at $T<U$ can be written in the form
\begin{equation}
B_{tT} = \frac{\mathbb{E}^\mathbb{P}[\pi_TH_T\,|\,\mathcal{F}_t]}{\pi_t}
\end{equation}
where $\{\pi_t\}$ is the pricing kernel. We consider the positive
martingale $\{M_{t}\}_{0\leq t<U}$ that satisfies
\begin{equation}\label{martingale2}
\rd M_t
=-\,\sigma_1\frac{U}{U-t}\,\mathbb{E}^\mathbb{P}[X_U\,|\,\xi_{tU}]\,M_t\,\rd
W_t,
\end{equation}
and introduce the pricing kernel $\{\pi_t\}$
given by
\begin{equation}
\pi_t = M_{t}\,f(t,\xi_{tU}).
\end{equation}
The dependence of the pricing kernel on $\{\xi_{tU}\}$ implies that
interest rates fluctuate due to the information flow in the market
about the likely value of the macroeconomic factor $X_U$ at time
$U$. Since the information processes are Markovian, the price of the
defaultable discount bond can be expressed by
\begin{equation}\label{Markovprop}
B_{tT} = \frac{\mathbb{E}^\mathbb{P}\left[M_{T}f(T,\xi_{TU})H_T\,\big|\,\xi_{tT},\xi_{tU}\right]}{M_{t}f(t,\xi_{tU})},
\end{equation}
where $H_T$ is the bond payoff at maturity $T$. We now suppose that
the payoff of the credit-risky bond is a function of $X_T$ and the
value of the information process associated with $X_U$ at the bond's
maturity $T$, that is
\begin{equation}
H_T = H\left(X_T, \xi_{TU}\right).
\end{equation}
Due to the independence property of the information processes, the
price of the credit-risky discount bond can be written as follows:
\begin{equation}\label{indep}
B_{tT}=\frac{\mathbb{E}^\mathbb{P}\left[\mathbb{E}^\mathbb{P}\left[M_{T}f(T,\xi_{TU})H(X_T,\xi_{TU})\,\big|\,\xi_{tT}\right]\,\big|\,
\xi_{tU}\right]}{M_{t}f(t,\xi_{tU})}.
\end{equation}
By applying the conditional form of Bayes formula, we change the
measure to the bridge measure $\B$ with respect to which the outer
expectation is taken:
\begin{equation}\label{DB-BP}
B_{tT}=\frac{\mathbb{E}^{\mathbb{B}}\left[\mathbb{E}^\mathbb{P}\left[f(T,\xi_{TU})H(X_T,\xi_{TU})\,\big|\,\xi_{tT}\right]
\,\big|\,\xi_{tU}\right]}{f(t,\xi_{tU})}.
\end{equation}
At this stage, we define a random variable $Y_{tT}$ by
\begin{equation}
Y_{tT} = \xi_{TU} - \frac{U-T}{U-t}\xi_{tU}.
\end{equation}
Since $\{\xi_{tU}\}$ is a Brownian bridge under $\mathbb{B}$, we
know that $Y_{tT}$ is a Gaussian random variable with zero mean and
variance
\begin{equation}
\textup{Var}^{\mathbb{B}}[Y_{tT}] = \frac{(T-t)(U-T)}{(U-t)}.
\end{equation}
Next we introduce a standard Gaussian random variable $Y$ and we
write $Y_{tT} = \nu_{tT}Y$, where $\nu^2_{tT} =
\textup{Var}^{\mathbb{B}}[Y_{tT}].$ We can now express the price of
the defaultable discount bond in terms of $Y$ as
\begin{equation}
B_{tT} =  \frac{\mathbb{E}^{\mathbb{B}}\left[\mathbb{E}^\mathbb{P}\left[f\left(T,\nu_{tT}Y + \frac{U-T}{U-t}\xi_{tU}\right)H\left(X_T,\nu_{tT}Y + \frac{U-T}{U-t}\xi_{tU}\right)\,\big|\,\xi_{tT}\right] \,\big|\,\xi_{tU}\right]}{f(t,\xi_{tU})}.
\end{equation}
Since $f(T,Y,\xi_{tU})$ in the numerator does not depend on
$\xi_{tT}$, we can write
\begin{align}
B_{tT} = \frac{\mathbb{E}^{\mathbb{B}}\left[f\left(T,\nu_{tT}Y + \frac{U-T}{U-t}\xi_{tU}\right)\mathbb{E}^\mathbb{P}\left[H\left(X_T,\nu_{tT}Y + \frac{U-T}{U-t}\xi_{tU}\right)\,\big|\,\xi_{tT}\right] \,\big|\, \xi_{tU}\right]}{f(t,\xi_{tU})}.
\end{align}
Because both $Y$ and $\xi_{tU}$ are independent of $\xi_{tT}$, the
inner conditional expectation in this expression can be carried out
explicitly. We obtain
\begin{equation}\label{Y-price}
B_{tT} = \frac{\mathbb{E}^{\mathbb{B}}\left[f\left(T,\nu_{tT}Y +
\frac{U-T}{U-t}\xi_{tU}\right)\sum_{i=0}^n
\pi_{it}\,H\left(x_i,\nu_{tT}Y +
\frac{U-T}{U-t}\xi_{tU}\right)\,\big|\,\xi_{tU}\right]}{f(t,\xi_{tU})},
\end{equation}
where $\pi_{it}$ denotes the conditional density of $X_T$, given by
\begin{equation}\label{cond-dens}
\pi_{it} = \mathbb{P}\left[X_T = x_i \,\big|\,\xi_{tT}\right] = \frac{p_i\exp{\left[\frac{T}{T-t}\left(\sigma_2x_i\xi_{tT}-\half\sigma^2_2x^2_it\right)\right]}}{\sum_{i=0}^n p_i\exp{\left[\frac{T}{T-t}\left(\sigma_2x_i\xi_{tT}-\half\sigma^2_2x^2_it\right)\right]}}.
\end{equation}
Since the random variable $\xi_{tU}$, appearing in the arguments of
$f(T,Y,\xi_{tU})$ and of $H(Y,\xi_{tU})$ in (\ref{Y-price}), is
measurable at time $t$ and $Y$ is independent of the conditioning
random variable $\xi_{tU}$, the conditional expectation reduces to a
Gaussian integral over the range of the random variable $Y$:
\begin{align} \label{credit risky bond}
B_{tT} = \frac{1}{f(t,\xi_{tU})}\sum_{i=0}^n \pi_{it} \int_{-\infty}^{\infty} &f\left(T,\nu_{tT}y + \frac{U-T}{U-t}\xi_{tU}\right) H\left(x_i,\nu_{tT}y + \frac{U-T}{U-t}\xi_{tU}\right)\nn\\
&\times \frac{1}{\sqrt{2\pi}}\exp{\left(-\half y^2\right)}\,\rd y.
\end{align}

In the case where the payoff is $H_T=X_T$, by using the expression
for the sovereign bond given by equation (\ref{hm}), we can write
the price of the defaultable bond as:
\begin{equation}
B_{tT} = P_{tT}\sum_{i=0}^n\pi_{it}\,x_i,
\end{equation}
where $\pi_{it}$ is defined by equation (\ref{cond-dens}). For
$n=1$, the defaultable bond pays a principal of $x_1$ units of
currency, if there is no default, and $x_0$ units of currency in the
event of default; we call such an instrument a ``binary bond''. In
particular, if $x_0=0$ and $x_1=1$, we call such a bond a ``digital
bond''. The price of the digital bond is
\begin{equation}\label{digi}
B_{tT} = P_{tT}\pi_{1t}.
\end{equation}

We can generalize the above situation slightly by considering a
pricing kernel $\{\pi_t\}$ of the form
\begin{equation}
\pi_t = M_t\,f(t, \xi_{tT}, \xi_{tU}).
\end{equation}
By following the technique in equations (\ref{Markovprop}) to
(\ref{credit risky bond}), and by using the fact that at time $T$ we
have $\xi_{TT} = \sigma_2X_TT$, we can show that
\begin{align}\label{generalized cr bond}
B_{tT} = \frac{1}{f(t,\xi_{tT}, \xi_{tU})}\sum_{i=0}^n \pi_{it} &\int_{-\infty}^{\infty}f\left(T, \sigma_2 x_i T, \nu_{tT}y + \frac{U-T}{U-t}\xi_{tU}\right)\nn\\
&\times\, H\left(x_i,\nu_{tT}y + \frac{U-T}{U-t}\xi_{tU}\right)\frac{1}{\sqrt{2\pi}}\exp{\left(-\half y^2\right)}\,\rd y.
\end{align}
Here we model the situation in which the pricing kernel in the
economy is not only a function of information at that time about the
macroeconomic variable, but is also dependent on noisy information
about potential default of the firm leaked in the market through
$\{\xi_{tT}\}$. This is relevant in light of events occurring in
financial markets where defaults by big companies can affect
interest rates and the market price of risk.

A measure for the excess return provided by a defaultable bond over
the return on a sovereign bond with the same maturity, is the bond
yield spread. This measure is given by the difference between the
yields-to-maturity on the defaultable bond and the sovereign bond,
see for example Bielecki \& Rutkowski (2002). That is:
\begin{equation}
s_{tT}=y^{d}_{tT} - y_{tT}
\end{equation}
for $t<T$, where $y_{tT}$ and $y^d_{tT}$ are the yields associated
with the sovereign bond and the credit-risky bond, respectively. We
have:
\begin{equation}
s_{tT}=\frac{1}{T-t}\left(\ln{P_{tT}} - \ln{B_{tT}}\right).
\end{equation}
In particular, the bond yield spread between a digital bond and the
sovereign bond is given by
\begin{equation}
s_{tT}=-\frac{1}{T-t}\ln{\pi_{1t}}.
\end{equation}
For bonds with payoff $H_T=X_T$, we see that the information related
to the macroeconomic factor $X_U$ does not influence the spread.
Thus for $0\leq t < T$, the spread at time $t$ depends only on the
information concerning potential default. In this case, the bond
yield spread between the defaultable discount bond and the sovereign
bond with stochastic interest rates is of the form of that in the
deterministic interest rate setting treated in Brody \emph{et al}.~(2007).

Figure 1 shows the bond yield spreads between a digital bond, with
all trajectories conditional on the outcome that the bond does not
default, and a sovereign bond. The maturities of the bonds are taken
to be $T=2$ years and the a priori probability of default is assumed to be $p_0=0.2$.
The effect of different values of the
information flow parameter is shown by considering $\sigma_2 =
0.04,\;\sigma_2 = 0.2$ and $\sigma_2 = 1,\; \sigma_2 =5$. Since the
paths of the digital bond are conditional on the outcome that
default does not occur, we observe that the bond yield spreads must
eventually drop to zero. The parameter $\sigma_2$ controls the
magnitude of genuine information about potential default that is
available to bondholders. For low values of $\sigma_2$, the
bondholder is, so to speak,``in the dark" about the outcome until
very close to maturity; while for higher values of $\sigma_2$, the
bondholder is better informed. As $\sigma_2$ increases, the
noisiness in the bond yield spreads, which is indicative of the
bondholder's uncertainty of the outcome, becomes less pronounced
near maturity. Furthermore, if the bondholders in the market were
well-informed, they would require a smaller premium for buying the
credit-risky bond since its behaviour would be similar to that of
the sovereign bond; this is illustrated in Figure 1. It is worth
noting that in the information-based asset pricing approach, an
increased level of genuine information available to investors about
their exposure, is manifestly equivalent to a sort of
``securitisation'' of the risky investments.

The case for which the paths of the digital bond are conditional on
default can also be simulated. Here, the effect of increasing the
information flow rate parameter $\sigma_2$ is similar. However, the
bondholder now requires an infinitely high reward for buying a bond
that will be worthless at maturity. Thus the bond-yield spread grows
to infinity at maturity.
\begin{figure}[H]
\begin{center}
\includegraphics[scale=0.44]{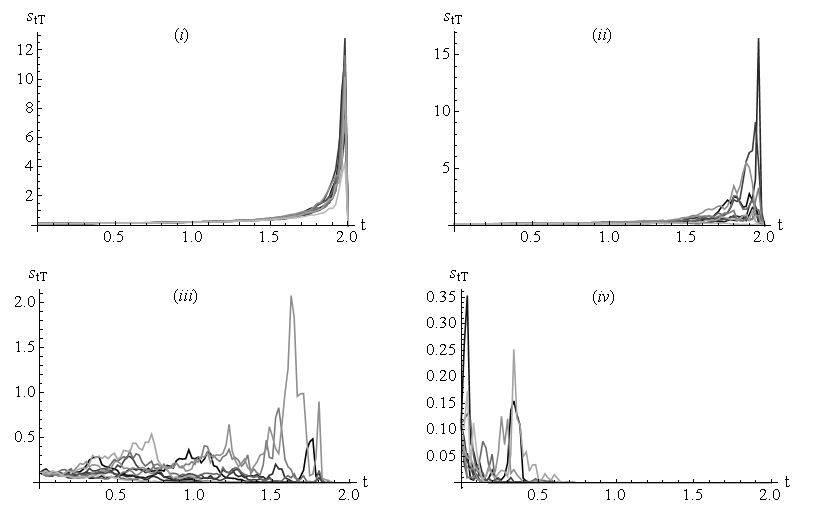}
\end{center}
\vspace{-0.9cm}
\caption{Bond yield spread between a digital bond (with all
trajectories conditional on no default) and a sovereign bond. The
bonds have maturity $T=2$ years. The a priori probability of default is taken to be $p_0=0.2$. We use (i) $\sigma_2=0.04$, (ii)
$\sigma_2=0.2$, (iii) $\sigma_2=1$, and (iv) $\sigma_2 = 5$.}
\end{figure}
\section{Credit-risky bonds with continuous market-\\dependent recovery}
Let us consider the case in which the credit-risky bond pays $H_T =
X_T$ where $X_T$ is a discrete random variable which takes values
$\{x_0, x_1, \ldots, x_n\} \in [0,1]$ with a priori probabilities
$\{p_0, p_1, \ldots, p_n\}$, where $x_n > x_{n-1} > \ldots > x_1 >
x_0$. Such a payoff spectrum is a model for random recovery where at
bond maturity one out of a discrete number of recovery levels may be
realised. We can also consider credit-risky bonds with continuous
random recovery in the event of default. In doing so, we introduce
the notion of ``market-dependent recovery''. Suppose that the payoff
of the defaultable bond is given by
\begin{equation}\label{H}
H_T = X_T + (1-X_T)\,R(\xi_{TU}),
\end{equation}
where $X_T$ takes the values $\{0,1\}$ with a priori
probabilities $\{p_0, p_1\}$. The recovery level $R : \mathbb{R}
\rightarrow [0,1)$ is dependent on the information at time $T$ about the macroeconomic factor $X_U$. In this case, if the credit-risky bond defaults at maturity $T$, the recovery level of the bond depends on the state of the economy at time $U$ that is
perceived in the market at time $T$. In other words, if the
sentiment in the market at time $T$ is that the economy will have
good times ahead, then a firm in a state of default at $T$ may have
better chances to raise more capital from liquidation (or
restructuring), thus increasing the level of recovery of the issued
bond. We can price the cash flow (\ref{H}) by applying equation
(\ref{credit risky bond}), with $n=1,\,x_0=0$ and $x_1=1$. The
result is:
\begin{align}\label{digbond-rec}
B_{tT} = P_{tT}\pi_{1t} + \pi_{0t}\,\frac{1}{f(t,\xi_{tU})}&\int_{-\infty}^{\infty}f\left(T, \nu_{tT}y + \frac{U-T}{U-t}\xi_{tU}\right)\nn\\
&\times R\left(\nu_{tT}y + \frac{U-T}{U-t}\xi_{tU}\right)
\frac{1}{\sqrt{2\pi}}\exp{\left(-\half y^2\right)}\,\rd y,
\end{align}
where $P_{tT}$ is given by equation (\ref{hm}). As an example,
suppose that we choose the recovery function to be of the form $R(z)
= 1-\exp{(-z^2)}.$ In this case, it is possible to have zero
recovery when the value of the information process at time $t$ is
$\xi_{tU}=-(U-t)/(U-T)\,\nu_{tT}Y$, thereby capturing the worst-case
scenario in which bondholders lose their entire investment in the
event of default.

The latter consideration is apt in the situation where the extent of
recovery is determined by how difficult it is for the firm to raise
capital by liquidating its assets, i.e. the exposure of the firm to
the general economic environment. However, this model does not say
much about how the quality of the management of the firm may
influence recovery in the event of default. This observation brings
us to another model of recovery. Default of a firm may be triggered
by poor internal practices and (or) tough economic conditions. We now
structure recovery by specifying the payoff of the credit risky bond
by
\begin{equation}\label{Recovery}
H_T = X_C\left[X_E + (1-X_E)R_E\right] + (1-X_C)\left[X_ER_C + (1-X_E)R_{CE}\right],
\end{equation}
where $X_C$ and $X_E$ are random variables taking values in
$\{0,1\}$ with a priori probabilities $\{p^{C}_0, p^{C}_1\}$ and
$\{p^E_0, p^E_1\}$, respectively. We define $X_C$ and $X_E$ to be
indicators of good management of the company and a strong economy,
respectively. We set $R_C$ to be a continuous random variable
assuming values in the interval $[0,1)$. We take $R_E$ to be a
function of $\xi_{TU}$, and $R_{CE}$ to be a function of $\xi_{TU}$
and $R_C$, where both $R_{E}$ and $R_{CE}$ assume values in the
interval $[0,1)$.

The payoff in equation (\ref{Recovery}) covers the following
situations: First, we suppose that despite good overall management
of the firm, default is triggered as a result of a depressed
economy. Here, $X_C = 1$ and $X_E=0$ which implies that $H_T = R_E$.
Therefore the recovery is dependent on the state of the economy at
time $T$ and thus, how difficult it has been for the firm to raise
funds. It is also possible that a firm can default in otherwise
favourable economic conditions, perhaps due to the management's
negligence. In this case we have $X_E=1$ and $X_C=0$. Thus $H_T=R_C$
and the amount recovered is dependent on the level of mismanagement
of the firm. Finally we have the worst case in which a firm is
poorly managed, $X_C=0$, and difficult economic times prevail,
$X_E=0$. Recovery is given by the amount $H_T = R_{CE}$, which is
dependent on both, the extent of mismanagement of the firm and how
much capital the firm can raise in the face of an economic downturn.
The particular payoff structure (\ref{Recovery}) is used in
Macrina (2006) to model the dependence structure between two
credit-risky discount bonds that share market factors in common.
Further investigation may include the situation where one models
such dependence structures for bonds subject to stochastic interest
rates and featuring recovery functions of the form (\ref{Recovery}).
\section{Call option price process}
Let $\{C_{st}\}_{0\le s\le t<T}$ be the price process of a
European-style call option with maturity $t$ and strike $K$, written
on a defaultable bond with price process $\{B_{tT}\}$. The price of
such an option at time $s$ is given by
\begin{equation}
C_{st}=\frac{1}{\pi_s}\,\E^{\PR}\left[\pi_t\left(B_{tT}-K\right)^+\vert\,\F_s\right].
\end{equation}
We recall that if the payoff of the credit-risky bond is $H_T=X_T$,
then the price of the bond at time $t$ is
\begin{equation}
B_{tT}=P_{tT}\sum^n_{i=0}\pi_{it}\,x_i,
\end{equation}
where $P_{tT}$ is given by equation (\ref{hm}) and the conditional
density $\pi_{it}$ is defined in equation (\ref{cond-dens}). The
filtration $\{\F_t\}$ is generated by the information
processes $\{\xi_{tT}\}$ and $\{\xi_{tU}\}$, and the pricing kernel
$\{\pi_{t}\}$ is of the form
\begin{equation}
\pi_t=M_t\,f(t,\xi_{tU}),
\end{equation}
with $\{M_t\}$ satisfying equation (\ref{martingale2}). Then the
price of the option at time $s$ is expressed by
\begin{equation}
C_{st}=\frac{1}{M_s\,f(s,\xi_{sU})}\,
\E^{\PR}\left[M_t\,f(t,\xi_{tU})\left(B_{tT}-K\right)^+\,\vert\,\xi_{sT},\xi_{sU}\right].
\end{equation}
We recall that the two information processes are independent, and
use the martingale $\{M_t\}$ to change the measure as follows:
\begin{equation}\label{Call5}
C_{st}=\frac{1}{f(s,\xi_{sU})}\,\E^{\B_U}\left[f(t,\xi_{tU})\,\E^{\PR}\left[\left(P_{tT}\sum^n_{i=0}\pi_{it}\,x_i-K\right)^+\bigg\vert\,\xi_{sT}\right]\bigg\vert\,\xi_{sU}\right].
\end{equation}
We first simplify the inner conditional expectation by following an
analogous calculation to that in Brody \emph{et al}.~(2007), Section 9. The
difference is that the discount factor $\{P_{tT}\}$ in
(\ref{Call5}) is stochastic. However since $\{P_{tT}\}$ is driven by
$\{\xi_{tU}\}$, it is unaffected by the conditioning of the inner
expectation, allowing us to use the result in Brody \emph{et al}.~(2007). Let us
introduce $\{\Phi_t\}$ by
\begin{equation}
\Phi_t=\sum^n_{i=0}p_{it},
\end{equation}
where $p_{it}=p_i\exp{\left[\frac{T}{T-t}\left(\sigma_2
x_i\,\xi_{tT}-\tfrac{1}{2}\,\sigma_2^2\,x_i^2\,t\right)\right]}$. We
write the inner expectation as
\begin{equation}
\E^{\PR}\left[\left(P_{tT}\sum^n_{i=0}\pi_{it}\,x_i-K\right)^+\bigg\vert\,\xi_{sT}\right]
=\E^{\PR}\left[\frac{1}{\Phi_t}\left(\sum^n_{i=0}\left(P_{tT}\,x_i-K\right)p_{it}\right)^+\bigg\vert\,\xi_{sT}\right].
\end{equation}
The process $\{\Phi_t^{-1}\}$ induces a change of measure from
$\PR$ to the bridge measure $\B_T$, under which $\{\xi_{tT}\}$ is a
Brownian bridge; this allows us to use Bayes formula to express the expectation
as follows:
\begin{align}
&\E^{\PR}\left[\frac{1}{\Phi_t}\left(\sum^n_{i=0}\left(P_{tT}\,x_i-K\right)p_{it}\right)^+\bigg\vert\,\xi_{sT}\right]
\nn\\
&\hspace{5.5cm}=\frac{1}{\Phi_s}\E^{\B_T}\left[\left(\sum^n_{i=0}\left(P_{tT}\,x_i-K\right)p_{it}\right)^+\bigg\vert\,\xi_{sT}\right].
\end{align}
In order to compute the expectation we introduce the Gaussian random
variable $Z_{st}$, defined by
\begin{equation}
Z_{st}=\frac{\xi_{tT}}{T-t}-\frac{\xi_{sT}}{T-s},
\end{equation}
which is independent of $\{\xi_{uT}\}_{0\le u\le s}$. It is possible
to find the critical value, for which the argument of the
expectation vanishes, in closed form if it is assumed that the
defaultable bond is binary. So, for $n=1$, the critical value
$z^{\ast}$ is given by
\begin{equation}
z^{\ast}=\frac{\ln\left[\frac{\pi_{0s}(K-x_0 P_{tT})}{\pi_{1s}(x_1
P_{tT}-K)}\right]+\frac{1}{2}\sigma_2^2\left(x_1^2-x_0^2\right)\alpha_{st}^2\,
T^2}{\sigma_2\,(x_1-x_0)\,\alpha_{st}\,T},
\end{equation}
where $\alpha_{st}^2 = \textup{Var}^{\mathbb{B}_{T}}[Z_{st}]$. The
computation of the expectation amounts to two Gaussian integrals
reducing to cumulative normal distribution functions, which we
denote by $N[x]$. We obtain the following:
\begin{align}
\E^{\PR}\left[\left(P_{tT}\sum^1_{i=0}\pi_{it}\,x_i-K\right)^+\bigg\vert\,\xi_{sT}\right]
=&\pi_{1s}(P_{tT}x_1-K)N[d^+_s]\nn\\
&-\pi_{0s}(K-P_{tT}x_0)N[d^-_s],
\end{align}
where
\begin{equation}
d^{\pm}_s=\frac{\ln\left[\frac{\pi_{1s}(x_1
P_{tT}-K)}{\pi_{0s}(K-x_0
P_{tT})}\right]\pm\frac{1}{2}\sigma_2^2\left(x_1-x_0\right)^2\alpha_{st}^2\,
T^2}{\sigma_2\,(x_1-x_0)\,\alpha_{st}\,T}.
\end{equation}
We can now insert this intermediate result into equation
(\ref{Call5}) for $n=1$; we have
\begin{align}\label{Call13}
&C_{st}=\frac{1}{f(s,\xi_{sU})}\,\E^{\B_U}\left[f(t,\xi_{tU})\left[
\pi_{1s}(P_{tT}x_1-K)N[d^+_s]\right.\right.\nn\\
&\hspace{5.5cm}\left.\left.-\pi_{0s}(K-P_{tT}x_0)N[d^-_s]\right]\vert\,\xi_{sU}\right].
\end{align}
We emphasize that $\{P_{tT}\}$ is given by a function
$P(t,T,\xi_{tU})$ and thus is affected by the conditioning with
respect to $\xi_{sU}$. To compute the expectation in equation
(\ref{Call13}), we use the same technique as in Section 4 and
introduce the Gaussian random variable $Y_{st}$, defined by
\begin{equation}
Y_{st}=\xi_{tU}-\frac{U-t}{U-s}\xi_{sU},
\end{equation}
with mean zero and variance
$\nu^2_{st}=\textrm{Var}^{\mathbb{B}_{U}}[Y_{st}]$. Thus, as shown
in the previous sections, the outer conditional expectation reduces
to a Gaussian integral:
\begin{align}\label{finalcall}
C_{st}=\frac{1}{f(s,\xi_{sU})}&\int_{-\infty}^{\infty}f\left(t,\nu_{st}y+\frac{U-t}{U-s}\,\xi_{sU}\right)\frac{1}{\sqrt{2\pi}}\exp\left(-\tfrac{1}{2}y^2\right)\nonumber\\
&\times\left[\pi_{1s}\left(P\left(t,T,\nu_{st}y+\frac{U-t}{U-s}\,\xi_{sU}\right)x_1-K\right)N[d^+_s(y)]\right.\nonumber\\
&\hspace{.5cm}\left.-\pi_{0s}\left(K-P\left(t,T,\nu_{st}y+\frac{U-t}{U-s}\,\xi_{sU}\right)x_0\right)N[d^-_s(y)]\right]\,\rd
y.
\end{align}
Therefore we obtain a semi-analytical pricing formula for a call
option on a defaultable bond in a stochastic interest rate setting.
The integral in equation (\ref{finalcall}) can be evaluated using
numerical methods once the function $f(t,x)$ is specified.
\section{Hybrid securities}
So far we have focused on the pricing of credit-risky bonds with
stochastic discounting. The formalism presented in the above
sections can also be applied to price other types of securities. In
particular, as an example of a hybrid security, we show how to price
an inflation-linked credit-risky discount bond. While such a
security has inherent credit risk, it offers bondholders protection
against inflation. This application also gives us the opportunity to
extend the thus far presented pricing models to the case where $n$
independent information processes are employed. We shall call such
models, ``multi-dimensional pricing models''.

In what follows, we consider three independent information
processes, $\{\xi_{tT}\}$, $\{\xi_{tU_1}\}$ and $\{\xi_{tU_2}\}$,
defined by
\begin{align}
&\xi_{tT}=\sigma\,t\,X_T+\beta_{tT},&   &\xi_{tU_1}=\sigma_1\,
t\,X_{U_1}+\beta_{tU_1},& &\xi_{tU_2}=\sigma_2\,
t\,X_{U_2}+\beta_{tU_2},&
\end{align}
where $0\le t\le T<U_1\le U_2$. The positive random variable $X_T$
is discrete, while $X_{U_1}$, $X_{U_2}$ are assumed to be
continuous. The market filtration $\{\F_t\}$ is generated jointly by
the three information processes. Let $\{C_{t}\}_{t\geq 0}$ be a
price level process, e.g., the process of the consumer price index.
The price $Q_{tT}$, at time $t$, of an inflation-linked discount
bond that pays $C_T$ units of a currency at maturity $T$, is
\begin{equation}
Q_{tT}=\frac{\E^{\PR}\left[\pi_T C_T\vert\,\F_t\right]}{\pi_t}.
\end{equation}
We now make use of the ``foreign exchange analogy" [see, e.g., Brigo
\& Mercurio (2006), Brody \emph{et al}.~(2008), Hinnerich (2008),
Hughston (1998), Mercurio (2005)] in which the nominal pricing
kernel $\{\pi_t\}$, and the real pricing kernel $\{\pi^R_t\}$, are
viewed as being associated with ``domestic" and ``foreign" economies
respectively, with the price level process $\{C_t\}$, acting as an
``exchange rate". The process $\{C_t\}$ is expressed by the
following ratio:
\begin{equation}\label{CPI}
C_t=\frac{\pi^R_t}{\pi_t}.
\end{equation}
For further details about the modelling of the real and the nominal
pricing kernels, and the pricing of inflation-linked assets, we
refer to Hughston \& Macrina (2009). In what follows, we make use of
the method proposed in Hughston \& Macrina (2009) to price an
example of an inflation-linked credit-risky discount bond (ILCR)
that, at maturity $T$, pays a cash flow $H_T=C_T
H(X_T,\xi_{TU_1},\xi_{TU_2})$. The price $H_{tT}$ at time $t\le T$
of such a bond is
\begin{equation}\label{Hyb1}
H_{tT}=\frac{1}{\pi_t}\E^{\PR}\left[\pi^R_{T}\,H(X_T,\xi_{TU_1},\xi_{TU_2})\,\big\vert\,\F_t\right],
\end{equation}
where we have used relation (\ref{CPI}). We choose to model the real
and the nominal pricing kernels by
\begin{align}\label{nom-real}
&\pi_t=M^{(1)}_t M^{(2)}_t\,f(t,\xi_{tU_1},\xi_{tU_2})&
&\textrm{and}& &\pi^R_t=M^{(1)}_t
M^{(2)}_t\,g(t,\xi_{tU_1},\xi_{tU_2}),&
\end{align}
where $f(t,x,y)$ and $g(t,x,y)$ are two functions of three
variables. The process $\{M^{(i)}_t\}_{0\leq t \leq T < U_i}$ for $i=1,2$ is a martingale that induces a change of measure to the bridge measure $\mathbb{B}_{i}$. We recall that the information process $\{\xi_{tU_i}\}$ has the law of a Brownian bridge under the measure $\B_{i}$. In order to work out the expectation in (\ref{Hyb1}) with the pricing kernel models introduced in (\ref{nom-real}), we can also define a process $\{M_t\}$ by
\begin{equation}
M_t = M^{(1)}_t M^{(2)}_t,
\end{equation}
where $0\leq t\leq T < U_1 \leq U_2$. Since the information
processes $\{\xi_{tU_1}\}$ and $\{\xi_{tU_2}\}$ are independent,
$\{M_t\}$ is itself an $(\{\mathcal{F}_t\}, \mathbb{P})$-martingale,
with $M_0 = 1$ and $\mathbb{E}^\mathbb{P}[M_t]=1$. Thus $\{M_t\}$
can be used to effect a change of measure from $\mathbb{P}$ to a
bridge measure $\mathbb{B}$, under which the random variables
$\xi_{tU_1}$ and $\xi_{tU_2}$ have the distribution of a Brownian
bridge for $0\leq t\leq T < U_1$. This can be verified as follows:
$\{\xi_{tU_1}\}$ is a Gaussian process with mean
\begin{equation}
\mathbb{E}^\mathbb{B}[\xi_{tU_1}] = \mathbb{E}^{\mathbb{B}_1}\left[\frac{M_t}{M_t^{(1)}}\xi_{tU_1}\right] = \mathbb{E}^{\mathbb{B}_1}\left[M^{(2)}_t\right]\,\mathbb{E}^{\mathbb{B}_1}[\xi_{tU_1}] = 0,
\end{equation}
due to the independence property of $\{\xi_{tU_1}\}$ and
$\{\xi_{tU_2}\}$. Furthermore, for $0\leq s\leq t \leq T < U_1$, the
covariance is given by
\begin{equation}
\mathbb{E}^\mathbb{B}[\xi_{sU_1}\xi_{tU_1}] = \mathbb{E}^{\mathbb{B}_1}\left[M^{(2)}_t\right]\,\mathbb{E}^{\mathbb{B}_1}[\xi_{sU_1}\xi_{tU_1}] = \mathbb{E}^\mathbb{P}[M_t]\, \mathbb{E}^{\mathbb{B}_1}[\xi_{sU_1}\xi_{tU_1}] = \frac{s(U_1-t)}{U_1}.
\end{equation}
The same can be shown for $\{\xi_{tU_2}\}$.

By use of $\{M_t\}$ and the Bayes formula, and the fact that
$\{\xi_{tT}\}$, $\{\xi_{tU_1}\}$ and $\{\xi_{tU_2}\}$  are
$\{\F_t\}$-Markov processes, equation (\ref{Hyb1}) reduces to
\begin{align}\label{master b-exp}
&H_{tT}=\nn\\
&\frac{1}{f(t,\xi_{tU_1},\xi_{tU_2})}
\E^{\B}\left[\E^{\PR}\left[g(T,\xi_{TU_1},\xi_{TU_2})H\left(X_T,\xi_{TU_1},\xi_{TU_2}\right)\,\big\vert\,\xi_{tT}\right]\big\vert\,\xi_{tU_1},\xi_{tU_2}\right].
\end{align}
Next we repeat an analogous calculation to the one leading from
equation (\ref{DB-BP}) to expression (\ref{credit risky bond}). For
the ILCR discount bond under consideration, we obtain
\begin{align}
H_{tT}=\frac{1}{f(t,\xi_{tU_1},\xi_{tU_2})}\sum^n_{i=0}\pi_{it}\int_{-\infty}^\infty\int_{-\infty}^{\infty}
&g\left(T,z(y_1),z(y_2)\right)H\left(x_i,z(y_1),z(y_2)\right)\nn\\
&\times\tfrac{1}{2\pi}\,\exp\left[-\tfrac{1}{2}\left(y^2_1+y^2_2\right)\right]\,\rd
y_1\,\rd y_2.
\end{align}
Here the conditional density $\pi_{it}$ is given by an expression analogous to the one in equation
(\ref{cond-dens}) and, $z(y_k)$ is defined for $k=1,2$ by
\begin{align}
z(y_k)=\nu^{(k)}_{tT}\,y_k+\frac{U_k-T}{U_k-t}\,\xi_{tU_k},\quad
\textrm{where}\quad
\nu^{(k)}_{tT}=\sqrt{\frac{(T-t)(U_k-T)}{U_k-t}}.&
\end{align}
In the special case where $H_T=X_T$, the expression for the price at
time $t$ of the ILCR discount bond simplifies to
\begin{equation}\label{Simple ILCR}
H_{tT}=Q_{tT}\sum^n_{i=0}\pi_{it}\,x_i.
\end{equation}
Here $Q_{tT}$ is the price of an inflation-linked discount bond that
depends on the information processes $\{\xi_{tU_1}\}$ and
$\{\xi_{tU_2}\}$. In particular, a formula similar to
(\ref{finalcall}) can be derived for the price of a European-style
call option written on an ILCR bond with price process given by
(\ref{Simple ILCR}) with $n=1$. We note here that similar pricing formulae can
be derived for credit-risky discount bonds traded in a foreign
currency. In this case the real pricing kernel, and thus the real
interest rate, is associated with the pricing kernel denominated in
the foreign currency. On the other hand, the nominal pricing kernel
is associated with the domestic currency, thus giving rise to the
domestic interest rate.
\section{Credit-risky coupon bonds}
Let $\{T_k\}_{k=1,\ldots,n}$ be a collection of fixed dates where
$0\le t\le T_1\le\ldots\le T_n$. We consider the valuation of a
credit-risky bond with coupon payment $H_{T_k}$ at time $T_k$ and
maturity $T_n$. The bond is in a state of default as soon as the
first coupon payment does not occur. We denote the price process of
the coupon bond by $\{B_{tT_n}\}$ and introduce $n$ independent
random variables $X_{T_1},\ldots,X_{T_n}$ that are applied to
construct the cash flows $H_{T_k}$ given by
\begin{align}
&H_{T_k}= \mathbf{c}\prod^k_{j=1}X_{T_j},
\end{align}
for $k=1,\ldots,n-1$, and for $k=n$ by
\begin{align}
H_{T_n}= \mathbf{(c+p)}\prod^n_{j=1}X_{T_j}.
\end{align}
Here $\mathbf{c}$ and $\mathbf{p}$ denote the coupon and principal payment,
respectively, and the random variables $\{X_{T_k}\}_{k=1,\ldots,n}$
take values in $\{0,1\}$. With each factor $X_{T_k}$ we associate an
information process $\{\xi_{tT_k}\}$ defined by
\begin{equation}
\xi_{tT_k} = \sigma_k\,t\,X_{T_k} + \beta_{tT_k}.
\end{equation}
Furthermore we introduce another information process
$\{\xi_{tU}\}$ given by
\begin{equation}
\xi_{tU} = \sigma\,t\,X_{U} + \beta_{tU}\qquad (0\le t\le T_n<U)
\end{equation}
that we reserve for the modelling of the pricing kernel. The market
filtration $\{\mathcal{F}_t\}$ is generated jointly by the $n+1$
information processes, that is $\{\xi_{tT_k}\}_{k=1,\ldots,n}$ and
$\{\xi_{tU}\}$. Following the method in Section 4, we model the
pricing kernel $\{\pi_t\}$ by
\begin{equation}
\pi_t = M_{t}\,f(t,\xi_{tU}),
\end{equation}
where the density martingale $\{M_t\}$ which induces a change of
measure to the bridge measure satisfies equation (\ref{martingale}).
Armed with these ingredients we are now in the position to write
down the formula for the price $B_{tT_n}$ at time $t$ of the
credit-risky coupon bond:
\begin{align}\label{coupon bond}
B_{tT_n}&=\frac{1}{\pi_t}\sum^n_{k=1}\,\E^{\PR}\left[\pi_{T_k}\,H_{T_k}\,\big\vert\,\xi_{tT_1},\ldots,
\xi_{tT_k}, \xi_{tU}\right],\\
&=\frac{1}{M_t\,f(t,\xi_{tU})}\sum^n_{k=1}\,\E^{\PR}\left[M_{T_k}\,f(T_k,\xi_{T_kU})\,\mathbf{c}\prod^{k}_{j=1}X_{T_j}\,\bigg\vert\,\xi_{tT_1},\ldots,
\xi_{tT_k}, \xi_{tU}\right]\nn\\
&\indent +\frac{1}{M_t\,f(t,\xi_{tU})}\,\E^{\PR}\left[M_{T_n}\,f(T_n,\xi_{T_nU})\,\mathbf{p}\prod^{n}_{j=1}X_{T_j}\,\bigg\vert\,\xi_{tT_1},\ldots,
\xi_{tT_n}, \xi_{tU}\right].&\nn
\end{align}
To compute the expectation, we use the approach presented in Section
4. Since the pricing kernel and the cash flow random variables
$H_{T_k}$, $k=1,\ldots,n$, are independent, we conclude that the
expression for the bond price $B_{tT_n}$ simplifies to
\begin{align}
B_{tT_n}=\mathbf{c}\,\sum^n_{k=1}P_{tT_k}&\E^{\PR}\left[\prod^k_{j=1}X_{T_j}\,\bigg\vert\,\xi_{tT_1},\ldots,\xi_{tT_k}\right]
\nn\\&+\mathbf{p}\,P_{tT_n}\E^{\PR}\left[\prod^n_{j=1}X_{T_j}\,\bigg\vert\,\xi_{tT_1},\ldots,\xi_{tT_n}\right],
\end{align}
where the discount bond system $\{P_{tT_k}\}$ is given by
\begin{equation}
P_{tT_k}=\frac{1}{f(t,\xi_{tU})}\int_{-\infty}^\infty f\left(T_k,\nu_{tT_k}y_k+\frac{U-T_k}{U-t}\,\xi_{tU}\right)\frac{1}{\sqrt{2\pi}}\exp\left(-\tfrac{1}{2}y_k^2\right)\rd
y_k,
\end{equation}
and $\nu^2_{tT_k}=(T_k-t)(U-T_k)/(U-t)$. We note that formula
(\ref{coupon bond}) can be simplified further since the expectations
therein can be worked out explicitly due to the independence
property of the information processes. We have,
\begin{equation}
\E^{\PR}\left[\prod^k_{j=1}X_{T_j}\,\bigg\vert\,\xi_{tT_1},\ldots,\xi_{tT_k}\right]
=\prod^k_{j=1}\pi_{1t}^{(j)},
\end{equation}
where the conditional density $\pi_{1t}^{(j)}$ at time $t$ that the
random variable $X_{T_j}$ takes value one is given by
\begin{equation}
\pi^{(j)}_{1t} = \frac{p^{(j)}_1
\exp{\left[\frac{T_j}{T_j-t}\left(\sigma_{j}\,\xi_{tT_j}-\half\sigma_{j}^2t\right)\right]}}{p^{(j)}_0+p^{(j)}_1
\exp{\left[\frac{T_j}{T_j-t}\left(\sigma_{j}\,\xi_{tT_j}-\half\sigma_{j}^2t\right)\right]}}.
\end{equation}
Here $p^{(j)}_1=\PR[X_{T_j}=1]$. Thus, the price $B_{tT_n}$ at time
$t$ of the credit-risky coupon bond is given by
\begin{equation}
B_{tT_n}=\sum_{k=1}^n\mathbf{c}\,P_{tT_k}\prod_{j=1}^{k}\pi_{1t}^{(j)}
+ \mathbf{p}\,P_{tT_n}\prod_{j=1}^{n}\pi^{(j)}_{1t}.
\end{equation}

At this stage, we observe that the price of a credit-risky coupon
bond has been derived for the case in which the cash flow functions
$H_{T_k},\;k=1,\ldots,n,$ do not depend on the information available
at time $T_k$ about the macroeconomic factor $X_U$, thereby leading
to independence between the discount bond system and the
credit-risky component of the bond. This is generalized in a
straightforward manner by considering cash flow functions of the
form
\begin{equation}
H_{T_k} = H(X_{T_1}, \ldots, X_{T_k}, \xi_{T_kU}),
\end{equation}
for $k=1,\ldots,n$. The valuation of such cash flows at time $t$ may
include the case treated in (\ref{credit risky
bond}), however endowed with coupon payments.

As an illustration we consider the situation in which the bond pays
a coupon $\mathbf{c}$ at $T_k,\; k=1,\ldots,n,$ and the principal
amount $\mathbf{p}$ at $T_n$. Upon default, market-dependent
recovery given by $R_k(\xi_{T_kU})$ (as a percentage of coupon plus
principal) is paid at $T_k$. For simplicity, we consider $n=2$. In
this case, the random cash flows of the bond are given by
\begin{align}
H_{T_1} &= \mathbf{c}X_{T_1}+ \mathbf{(c+p)}\,R_1(\xi_{T_1U})(1-X_{T_1}),\nn\\
H_{T_2} &= \mathbf{(c+p)}\,X_{T_1}\left[X_{T_2}+R_2(\xi_{T_2U})(1-X_{T_2})\right].\nn
\end{align}
By making use of the technique presented in Section 5, we can
express the price of the credit-risky coupon bond by
\begin{align*}
B_{tT_2}&=\mathbf{c}\,P_{tT_1}\pi^{(1)}_{1t} + \mathbf{(c+p)}\,P_{tT_2}\pi^{(1)}_{1t}\pi^{(2)}_{1t}\nn\\
&+\mathbf{(c+p)}\left[\pi^{(1)}_{0\,t} \frac{1}{f(t,\xi_{tU})} \int_{-\infty}^\infty f\left(T_1, m(y_1)\right)\,R_1\left(m(y_1)\right)\frac{1}{\sqrt{2\pi}}\exp{\left(-\tfrac{1}{2}y^2_1\right)}\,\rd y_1\right.\nn\\
&\hspace{.5cm}\left.+\pi^{(1)}_{1t}\pi^{(2)}_{0\,t}\frac{1}{f(t,\xi_{tU})}
\int_{-\infty}^\infty f\left(T_2,
m(y_2)\right)\,R_2\left(m(y_2)\right)\frac{1}{\sqrt{2\pi}}\exp{\left(-\tfrac{1}{2}y^2_2\right)}\,\rd
y_2\right],
\end{align*}
where, for $k=1,2$, we define
\begin{align}
m(y_k)=\nu_{tT_k}\,y_k+\frac{U-T_k}{U-t}\,\xi_{tU},\quad \quad
\nu_{tT_k}=\sqrt{\frac{(T_k-t)(U-T_k)}{U-t}}.&
\end{align}
\section{Credit-sensitive pricing kernels}
We fix the dates $T_1$ and $T_2$, where $T_1\leq T_2$, to which we
associate the economic factors $X_{T_1}$ and $X_{T_2}$ respectively.
The first factor is identified with a debt payment at time $T_1$.
For example $X_{T_1}$ could be a coupon payment that a country is
obliged to make at time $T_1$. The second factor, $X_{T_2}$, could
be identified with the measured growth (possibly negative) in the
employment level in the same country at time $T_2$ since the last
published figure. In such an economy, with two random factors only,
it is plausible that the prices of the treasuries fluctuate
according to the noisy information market participants will have
about the outcome of $X_{T_1}$ and $X_{T_2}$. Thus the price of a
sovereign bond with maturity $T$, where $0\le t\le T<T_1\leq T_2$, is
given by:
\begin{align}
P_{tT}=\frac{1}{f(t,\xi_{tT_1},\xi_{tT_2})}\int^{\infty}_{-\infty}\int^{\infty}_{-\infty}&f\left(T,\nu^{(1)}_{tT}y_1+\frac{T_1-T}{T_1-t}\xi_{tT_1},\nu^{(2)}_{tT}y_2+\frac{T_2-T}{T_2-t}\xi_{tT_2}\right)\nn\\
&\times \frac{1}{2\pi}\exp\left[-\tfrac{1}{2}\left(y^2_1+y^2_2\right)\right]\,\rd
y_2\,\rd y_1.
\end{align}
In particular, the resulting interest rate process in this model is
subject to the information processes $\{\xi_{tT_1}\}$ and
$\{\xi_{tT_2}\}$ making it fluctuate according to information (both
genuine and misleading) about the economy's factors $X_{T_1}$ and
$X_{T_2}$.\\

We now ask the following question: {\it What type of model should
one consider if the goal is to model a pricing kernel that is
sensitive to an accumulation of losses}? {\it Or in other words,
how should one model the nominal short rate of interest and the
market price of risk processes if both react to the amount of debt
accumulated by a country over a finite period of time}?\\
\\
To treat this question we need to introduce a model for an
accumulation process. We shall adopt the method developed in
Brody \emph{et al}.~(2008b), where the idea of a gamma bridge information process is
introduced. It turns out that the use of such a cumulative process is
suitable to provide an answer to the question above. In fact, if in
the example above, the factor $X_{T_1}$ is identified with the total
accumulated debt at time $T_1$, then the gamma bridge information
process $\{\xi^{\gamma}_{tT_1}\}$, defined by
\begin{equation}
\xi^{\gamma}_{tT_1}=X_{T_1}\,\gamma_{tT_1}
\end{equation}
where $\{\gamma_{tT_1}\}_{0\le t\le T_1}$ is a gamma bridge process
that is independent of $X_{T_1}$, measures the level of the
accumulated debt as of time $t$, $0\le t\le T_1$. If the market
filtration is generated, among other information processes, also by
the debt accumulation process, then asset prices that are calculated
by use of this filtration, will fluctuate according to the updated
information about the level of the accumulated debt of a country. We
now work out the price of a sovereign bond for which the
price process reacts both to Brownian and gamma information.

We consider the time line $0\le t\le T< T_1\leq T_2<\infty$. Time $T$
is the maturity date of a sovereign bond with unit payoff and price
process $\{P_{tT}\}_{0\le t\le T}$. With the date $T_1$ we associate
the factor $X_{T_1}$ and with the date $T_2$ the factor $X_{T_2}$.
The positive random variable $X_{T_1}$ is independent of $X_{T_2}$,
and both may be discrete or continuous random variables. Then we
introduce the following information processes:
\begin{align}
&\xi^{\gamma}_{tT_1}=X_{T_1}\,\gamma_{tT_1}, & \xi_{tT_2}=\sigma\,t
\,X_{T_2} +\beta_{tT_2}.
\end{align}
The process $\{\xi^{\gamma}_{tT_1}\}$ is a gamma
bridge information process, and it is taken to be independent of
$\{\xi_{tT_2}\}$. The properties of the gamma bridge process
$\{\gamma_{tT_1}\}$ are described in great detail in Brody \emph{et
al}.~(2008b). We assume that the market filtration $\{\F_t\}_{t\geq
0}$ is generated jointly by $\{\xi^{\gamma}_{tT_1}\}$ and
$\{\xi_{tT_2}\}$.

In this setting, the pricing kernel reacts to the updated information
about the level of accumulated debt and, for the sake of example,
also to noisy information about the likely level of employment
growth at $T_2$. Thus we propose the following model for the pricing kernel:
\begin{equation}\label{cs-pk}
\pi_t=M_{t}\,f\left(t,\xi^{\gamma}_{tT_1},\xi_{tT_2}\right)
\end{equation}
where the process $\{M_t\}$ is the change-of-measure martingale from the probability
measure $\PR$ to the Brownian bridge measure $\mathbb{B}$, satisfying
\begin{equation}
\rd
M_{t}=-\,\sigma\frac{T_2}{T_2-t}\,\E\left[X_{T_2}\,\vert\,\xi_{tT_2}\right]M_{t}\,\rd
W_t.
\end{equation}
Here $\{W_t\}$ is an $(\{\F_t\}, \mathbb{P})$-Brownian motion. The formula for the
price of the sovereign bond is given by
\begin{equation}
P_{tT}=\frac{\E^\PR\left[M_{T}f\left(T,\xi^{\gamma}_{TT_1},\xi_{TT_2}\right)\big\vert\,\xi^{\gamma}_{tT_1},\xi_{tT_2}\right]}
{M_{t}\,f\left(t,\xi^{\gamma}_{tT_1},\xi_{tT_2}\right)}.
\end{equation}
We make use of the Markov property and the independence
property of the information processes, together with the change of
measure to express the bond price by
\begin{eqnarray}
P_{tT}=\frac{\E^\PR_{\gamma}\left[\E^{\B}\left[f\left(T,\xi^{\gamma}_{TT_1},\xi_{TT_2}\right)\big\vert\,\xi_{tT_2}\right]\big\vert\,\xi^{\gamma}_{tT_1}\right]}{f\left(t,\xi^{\gamma}_{tT_1},\xi_{tT_2}\right)}.
\end{eqnarray}
Here, the expectations $\E^\PR_{\gamma}$ and $\E^\mathbb{B}$ are operators
that apply according to the dependence of their argument on the
random variables $\xi^{\gamma}_{TT_1}$ and $\xi_{TT_2}$
respectively. This is a direct consequence of the independence of
$\{\xi^{\gamma}_{tT_1}\}$ and $\{\xi_{tT_2}\}$. We now use the
technique adopted in the preceding sections, where we introduce the
Gaussian random variable $Y_{tT}$ with mean zero and variance $\nu^2_{tT} = (T-t)(T_2-T)/(T_2-t),$ and the standard
Gaussian random variable $Y$. By following the approach taken in
Section 4, we can compute the inner expectation explicitly since the
conditional expectation reduces to a Gaussian integral over the
range of the random variable $Y$. Thus we obtain:
\begin{equation}\label{bond-betafinal}
P_{tT}=\int^{\infty}_{-\infty}\frac{\E^\PR_{\gamma}\left[f\left(T,\xi^{\gamma}_{TT_1},\nu_{tT}y+\frac{T_2-T}{T_2-t}\xi_{tT_2}\right)\,\big\vert\,\xi^{\gamma}_{tT_1}\right]}{f\left(t,\xi^{\gamma}_{tT_1},\xi_{tT_2}\right)}\,\frac{1}{\sqrt{2\pi}}\exp\left(-\tfrac{1}{2}y^2\right)\,\rd
y.
\end{equation}
The feature of this model which sets it apart from those considered
in preceding sections, is the fact that we have to calculate a gamma
expectation $\E^\PR_{\gamma}$. In this case, we cannot adopt the
``usual" change-of-measure method we have used thus far. To this end
we refer to the work in Brody \emph{et al}.~(2008b), where the price
process of the Arrow-Debreu security for the case that it is driven
by a gamma bridge information process is derived. We use this result
and obtain for the Arrow-Debreu density process $\{A_{tT}\}$ the
following expression:
\begin{align}\label{ADPP-formula}
&A_{tT}(y_{\gamma})={\mathbb E}^\PR\left[\delta(\xi^{\gamma}_{TT_1}-y_{\gamma})\,\big\vert\,\xi^{\gamma}_{tT_1}\right]\\
&=\frac{{\indi
1}{\{y_{\gamma}>\xi^{\gamma}_{tT_1}\}}\,(y_{\gamma}-\xi^{\gamma}_{tT_1})^{m(T-t)-1}}
{B[m(T-t),m(T_1-T)]}\frac{\int^{\infty}_{y_{\gamma}}p(x)\,
x^{1-mT_1}(x-y_{\gamma})^{m(T_1-T)-1}\rd
x}{\int^{\infty}_{\xi^{\gamma}_{tT_1}}
p(z)\,z^{1-mT_1}(z-\xi^{\gamma}_{tT_1})^{m(T_1-t)-1} \rd
z},\label{ADPP-explicit}
\end{align}
where $\delta(y)$ is the Dirac distribution and $p(x)$ is the a
priori probability density of $X_{T_1}$. Here $B[a,b]$ is the beta
function. Following Macrina (2006), Section 3.4, we consider a
function $h(\xi^{\gamma}_{TT_1})$ of the random variable
$\xi^{\gamma}_{TT_1}$ and note that for a suitable function $h$ we
may write:
\begin{equation}
\E^\PR_{\gamma}\left[h\left(\xi^{\gamma}_{TT_1}\right)\,\big\vert\,\xi^{\gamma}_{tT_1}\right] =\int^{\infty}_{-\infty}\E^\PR_{\gamma}\left[\delta\left(\xi^{\gamma}_{TT_1}-y_{\gamma}\right)\big\vert\,\xi^{\gamma}_{tT_1}\right]h(y_{\gamma})\,\rd
y_{\gamma}.\label{Gen-funct-AD}
\end{equation}
Next we see that the conditional expectation under the integral is
the Arrow-Debreu density (\ref{ADPP-formula}) for which there is the
closed-form expression (\ref{ADPP-explicit}). We go back to equation
(\ref{bond-betafinal}) and observe that the conditional expectation
under the integral is of the form
$\E^\PR_{\gamma}\left[h\left(\xi^{\gamma}_{TT_1}\right)\big\vert\xi^{\gamma}_{tT_1}\right]$.
Thus we can use (\ref{Gen-funct-AD}) to calculate the gamma
expectation in (\ref{bond-betafinal}). We write:
\begin{align}\label{f-AD}
&\E^\PR_{\gamma}\left[f\left(T,\xi^{\gamma}_{TT_1},\nu_{tT}\,y+\frac{T_2-T}{T_2-t}\,\xi_{tT_2}\right)\Big\vert\,\xi^{\gamma}_{tT_1}\right]\nn\\
&\hspace{3cm}=\int^{\infty}_{-\infty}A_{tT}\left(y_{\gamma}\right)f\left(T,y_{\gamma},\nu_{tT}\,y + \frac{T_2-T}{T_2-t}\,\xi_{tT_2}\right)\rd
y_{\gamma}.
\end{align}
We are now in the position to write down the bond price
(\ref{bond-betafinal}) in explicit form by using equation
(\ref{f-AD}). We thus obtain:
\begin{equation}
P_{tT}=\int^{\infty}_{-\infty}\int^{\infty}_{-\infty}\,\frac{A_{tT}\left(y_{\gamma}\right)f\left(T,y_{\gamma},\nu_{tT}\,y + \frac{T_2-T}{T_2-t}\,\xi_{tT_2}\right)}
{f\left(t,\xi^{\gamma}_{tT_1},\xi_{tT_2}\right)}\,\frac{1}{\sqrt{2\pi}}\exp\left(-\tfrac{1}{2}y^2\right)\rd
y_{\gamma}\,\rd y.
\end{equation}
The bond price can be written more concisely by defining
\begin{align}
\tilde{f}\left(T,t,\xi^{\gamma}_{tT_1},\xi_{tT_2}\right) =
\int^{\infty}_{-\infty}\int^{\infty}_{-\infty}\,&A_{tT}\left(y_{\gamma}\right)f\left(T,y_\gamma,\nu_{tT}y + \frac{T_2-T}{T_2-t}\,\xi_{tT_2}\right)\nn\\
&\times \frac{1}{\sqrt{2\pi}}\exp\left(-\tfrac{1}{2}\,y^2\right)\rd y_{\gamma}\,\rd
y.
\end{align}
We thus have:
\begin{equation}
P_{tT}=\frac{\tilde{f}\left(T,t,\xi^{\gamma}_{tT_1},\xi_{tT_2}\right)}{f\left(t,\xi^{\gamma}_{tT_1},\xi_{tT_2}\right)}.
\end{equation}
Future investigation in this line of research incorporates the
constructions of processes $\{f(t,\xi^{\gamma}_{tT_1},\xi_{tT_2})\}$
such that the resulting pricing kernel (\ref{cs-pk}) is an
$(\{\mathcal{F}_t\}, \mathbb{P})$-supermartingale. The appropriate
choice of $f(t,x,y)$ depends also on a suitable description of the
economic interplay of the information flows modelled by
$\{\xi^{\gamma}_{tT_1}\}$ and $\{\xi_{tT_2}\}$. One might begin with
looking at the situation in which the price of the bond depreciates
due to a rising debt level and a higher level of employment. We
conclude by observing that the gamma bridge information process may
also be considered for the modelling of credit-risky bonds, where
default is triggered by the firm's accumulated debt exceeding a
specified threshold at bond maturity. Random recovery models may be
constructed using the technique in Section 5.
\section*{Acknowledgments}
The authors thank D. C. Brody, M. H. A. Davis, C. Hara,
T. Honda, E. Hoyle, R. Miura, H. Nakagawa, K. Ohashi, J. Sekine,
K. Tanaka, and participants in the KIER-TMU 2009 International Workshop
on Financial Engineering, Tokyo, and the seminars at Ritsumeikan University, Kusatsu,
and ICS Hitotsubashi University, Tokyo for useful comments. We are in particular grateful to J. Akahori and L. P.
Hughston for helpful suggestions at an early stage of this work. P.
A. Parbhoo thanks the Institute of Economic Research, Kyoto
University, for its hospitality, and acknowledges financial support
from the Programme in Advanced Mathematics of Finance at the University
of the Witwatersrand and the National Research Foundation, South Africa.

\section*{References}
\begin{enumerate}

\item Akahori, J., Hishida, Y., Teichmann, J. and Tsuchiya, T. (2009),``A Heat Kernel Approach to Interest Rate Models'', arXiv:0910.5033.

\item Akahori, J. and Macrina, A. (2010), ``Heat Kernel Interest Rate Models with Time-Inhomogeneous Markov
Processes'', Ritsumeikan University, King's College London and Kyoto
University working paper.

\item Bielecki, T. R. and Rutkowski, M. (2002), {\it Credit Risk: Modelling, Valuation and Hedging}, Springer-Verlag, Berlin.

\item Brigo, D. and Mercurio, F. (2006), {\it Interest Rate Models: Theory and Practice (with Smile, Inflation and Credit)}, Springer-Verlag, Berlin.

\item Brody, D. C., Crosby, J. and Li, H. (2008), ``Convexity Adjustments in Inflation-Linked Derivatives'', {\it Risk Magazine}, September issue, 124-129.

\item Brody, D. C., Davis, M. H. A., Friedman, R. L. and Hughston, L. P. (2009), ``Informed Traders'', {\it Proceedings of the Royal Society London}, {\bf A465}, 1103--1122.

\item Brody, D. C., Hughston, L. P. and Macrina, A. (2007), ``Beyond Hazard Rates: A New Framework to Credit Risk Modelling'', in \textit{Advances in Mathematical Finance, Festschrift Volume in Honour of Dilip Madan} (eds Elliott R., Fu, M., Jarrow, R. and Yen, J. Y.), Birkh\"auser, Basel.

\item Brody, D. C., Hughston, L. P. and Macrina, A. (2008a), ``Information-Based Asset Pricing'', {\it International Journal of Theoretical and Applied Finance}, {\bf 11}, 107--142.

\item Brody, D. C., Hughston, L. P. and Macrina, A. (2008b), ``Dam Rain and Cumulative Gain'', {\it Proceedings of the Royal Society London}, {\bf A464}, 1801--1822.

\item Flesaker, B. and Hughston, L. P. (1996), ``Positive Interest'', {\it Risk}, {\bf 9}, 46--49.

\item Hinnerich, M. (2008), ``Inflation-Indexed Swaps and Swaptions'', {\it Journal of Banking and Finance}, {\bf 32}, 2293--2306.

\item Hoyle, E., Hughston, L. P. and Macrina, A. (2009), ``L\'evy Random Bridges and the Modelling of Financial Information'', arXiv:0912.3652.

\item Hughston, L. P. (1998), ``Inflation Derivatives'', Merrill Lynch and King's College London working paper, with added note (2004).

\item Hughston, L. P. and Macrina, A. (2009), ``Pricing Fixed-Income Securities in an Information-Based Framework'', arXiv:0911.1610.

\item Hunt, P. J. and Kennedy, J. E. (2004), {\it Financial Derivatives in Theory and Practice}, Wiley, Chichester.

\item Macrina, A. (2006), ``An Information-Based Framework for Asset Pricing: $X$-factor Theory and its Applications'', PhD thesis, King's College London.

\item Mercurio, F. (2005), ``Pricing Inflation-Indexed Derivatives'', {\it Quantitative Finance}, {\bf 5}, No. 3, 289-302.

\item Rogers, L. C. G. (1997), ``The Potential Approach to the Term Structure of Interest Rates and Foreign Exchange Rates'', {\it Mathematical Finance}, {\bf 7}, 157--176.

\item Rutkowski, M. and Yu, N. (2007), ``An Extension of the Brody-Hughston-Macrina Approach to Modelling of Defaultable Bonds'', {\it International Journal of Theoretical and Applied Finance}, {\bf 10}, 557--589.

\end{enumerate}

\end{document}